\documentclass[10pt,conference]{IEEEtran}

\usepackage{tabularx,booktabs}
\usepackage{graphicx}
\usepackage{ragged2e}
\usepackage{lipsum} 
\usepackage{gensymb}
\usepackage{url}
\usepackage{float}

\newcommand{\etal}{\emph{et~al}.}
\newcommand{\ie}{\emph{i}.\emph{e}., }
\newcommand{\eg}{\emph{e}.\emph{g}., }
\newcommand{\cf}{\emph{cf}. }

\newcommand{\pattern}[1]{\textsc{\MakeLowercase{#1}}}

\usepackage{amsmath,amssymb}
\usepackage[ruled,vlined,linesnumbered]{algorithm2e}

\SetKwInOut{Input}{Input}
\SetKwInOut{Output}{Output}
\SetKwBlock{Init}{init}{end}
\SetKwComment{Comment}{// }{}
\SetCommentSty{itshape}
\SetKwBlock{OnInput}{onInput}{end}
\SetKwBlock{OnMax}{}{end}
\SetKwBlock{OnTimeout}{onTimeout}{end}
\SetKw{Trigger}{trigger}
\DontPrintSemicolon
\SetAlFnt{\small}

\begin{document}

\title{Empowering Visual Internet-of-Things Mashups with Self-Healing Capabilities}

\author{\IEEEauthorblockN{João Pedro Dias}
\IEEEauthorblockA{DEI, Faculty of Engineering,\\
University of Porto, and\\
INESC TEC\\
Email: jpmdias@fe.up.pt}
\and
\IEEEauthorblockN{André Restivo}
\IEEEauthorblockA{DEI, Faculty of Engineering,
\\University of Porto, and\\
LIACC\\
Email: arestivo@fe.up.pt}
\and
\IEEEauthorblockN{Hugo Sereno Ferreira}
\IEEEauthorblockA{DEI, Faculty of Engineering,\\
University of Porto, and\\
INESC TEC\\
Email: hugo.sereno@fe.up.pt}
}

\maketitle

\begin{abstract}
Internet-of-Things (IoT) systems have spread among different application domains, from home automation to industrial manufacturing processes. The rushed development by competing vendors to meet the market demand of IoT solutions, the lack of interoperability standards, and the overall lack of a defined set of best practices have resulted in a highly complex, heterogeneous, and frangible ecosystem. Several works have been pushing towards visual programming solutions to abstract the underlying complexity and help humans reason about it. As these solutions begin to meet widespread adoption, their building blocks usually do not consider reliability issues. Node-RED, being one of the most popular tools, also lacks such mechanisms, either built-in or via extensions. In this work we present SHEN (Self-Healing Extensions for Node-RED) which provides 17 nodes that collectively enable the implementation of self-healing strategies within this visual framework. We proceed to demonstrate the feasibility and effectiveness of the approach using real devices and fault injection techniques.
\end{abstract}

\IEEEpeerreviewmaketitle


\section{Introduction}

The pervasiveness of the Internet-of-Things (IoT) is reshaping how people interact with everyday objects, as connected devices profoundly influence our surroundings, in many ways and in an unprecedented fashion. The preponderance of vendor-specific devices and applications, along with their heterogeneity, in terms of protocols and standards, altogether with a large number of components (\eg devices and services) make these systems infeasible to deploy manually, setup, manage and maintain each component, thus exceeding the human ability to manage all connected devices~\cite{Tahir2019}. Such factors lead to the birth of several vendor and protocol-independent integration and orchestration solutions. These solutions attempt to abstract the IoT systems' underlying complexity mostly resulting from its heterogeneity and distributed nature while favoring low-code approaches that make it feasible for less technical users to develop and maintain their own IoT systems, with one of the most common solutions being Node-RED~\cite{noderedwebsite}.

Node-RED allows users to program their systems by connecting multiple data sources and actuators, and to define some logic flows, in a \textit{drag-n-drop} fashion. Although popular, it has several limitations, including: no proper mechanisms for debugging and testing \emph{flows} (existent works~\cite{SEDES2018,Ancona2018,Clerissi18,DiogoTorres2020} both identify and address these issues at some extent), no type-checking mechanism exists, the used visual abstractions are leaky~\cite{spolsky2004law} and there is no formal meta-model~\cite{prehofer2015}, it is designed to work in a centralized fashion and each particular component's behaviour (nodes) is mostly opaque. While most of these limitations can be bypassed by leveraging the \texttt{Function} node which runs any JavaScript code, this jeopardizes the goal of being a visual and low-code approach. 

Approaches such as Node-RED and others~\cite{BabaCheikh2020} attempt to ease the integration, development, and evolution of IoT systems requiring different technical expertise levels, mostly dismiss considerations regarding fault-tolerance and system reliability. This is common symptom across IoT due to its complexity; as stated by Javed \etal~\cite{Javed18}, \emph{``building a fault-tolerant system for IoT is a complex task, mainly because of the extremely large variety of edge devices, data computing technologies, networks, and other resources that may be involved in the development process''}.

However, as these devices and systems permeate our daily lives, their correct function becomes ever-more paramount. Some authors argue that most IoT device's failures are typical \emph{fail-stop}, \ie consistent, easily detectable, reproducible (easy to \textit{debug} and correct), and easy to fix by end-users by replacing the faulty unit~\cite{Terry2016}. Yet, \textit{fail-stop} failures can still impact the quality-of-life of users (\eg a broken smoke-detection device). Failures beyond those, including \textit{intermittent faults}, can have nefarious side-effects when fail-over options, anomaly detection and correction mechanisms are not in-place~\cite{Smith2017,Terry2016}.  

As a motivational example, consider a thermostat that stops working (\ie fail-stop), an AC unit can fallback to a predefined working temperature or shut down entirely. However, if the thermostat malfunctions so that it reports high-temperature readings, it can make the AC force the ambient temperature fall below unsafe levels for a newborn. Similarly, if a refrigerator's temperature sensor reports erroneous readings, it may cause food to degrade faster, possibly leading to food poisoning for the entire household~\cite{Abad2019}. 
Although most IoT systems, such as the smart app we use to control our home lights, are not considered safety-critical (working in a \textit{fire-and-forget} fashion), their malfunction can still cause discomfort or be life-threatening. While some device's proneness to failure can be reduced by improving their hardware parts or by adding redundancy (\eg triple modular redundancy and majority consensus), it typically comes with additional costs and complexity~\cite{Terry2016,Ni2009,Zhou15thesis}. 


Early research on the complexity of systems by IBM Research resulted in the birth of the concept of autonomic computing as a way of coping with the continuous growth in the complexity of operating, managing, and integrating computing systems~\cite{Psaier2011,Ganek2003}. Autonomic computing systems need to know and understand themselves to be able to adapt to ever-changing conditions autonomously. A system to be autonomic should have the following self-* properties: (1)~self-configuration, the ability of a system to automatically and seamlessly configure itself by following a set of high-level policies, (2)~self-optimization, the ability of a system to improve --- or maintain --- its performance without human intervention, (3)~self-protection, the ability of a system to protect itself from malicious attacks, and (4)~self-healing, the ability of a system to automatically detect, diagnosis and repair system defections at both hardware and software levels~\cite{Murch2004,Psaier2011}. Several authors~\cite{ashraf2015acacos,Angarita2015,vermesan2011,Aktas2019} have been adopting autonomic computing as a way to cope with the growing complexity of IoT systems and their cross-domain application.


In a previous study on adding self-healing capabilities to Node-RED in~\cite{selfheal20}, we provided a proof-of-concept focused on three concrete scenarios: (1)~the unavailability of the IoT system message broker, (2)~dealing with some erroneous sensor readings, and (3)~checking for connectivity issues. These were implemented using a mix of sub-\textit{flows}, \texttt{Function} nodes and new nodes. Preliminary tests were performed on a laboratory setup, which provided insights on the pending challenges that needed to be addressed to achieve proper self-healing within Node-RED, without modifying it. 

In this paper, we present a set of self-healing building Node-RED \textit{nodes}, that enable users to improve the resilience of IoT systems. We validated our approach on a physical testbed~\cite{europlop19}, which we named SmartLab, comprised of multiple heterogeneous sensors and actuators. A set of 3 scenarios showcase some of the cases in which such \textit{nodes} would provide benefits to existent Node-RED \textit{flows}, improving the overall capability of the system to withhold failures of its parts.

The remaining of the paper is structured as follows: Section~\ref{sec:related} summarizes relevant literature that focus similar issues, Section~\ref{sec:work} present an overview to our approach, Section~\ref{sec:experiments} summarizes the experiments and our observations. Section~\ref{sec:conclude} presents some final remarks and point to future work.

\section{Related Work}
\label{sec:related}

IoT systems have been primarily identified as a core example of a system that must contemplate autonomic components~\cite{ashraf2015acacos,Angarita2015,vermesan2011}. These components --- that can range from single devices (\eg smart locks) to whole systems (\eg smart homes) --- should be capable of self-management, reducing the need for frequent human operation~\cite{Kopetz2011}. This issue becomes even more important in critical systems and when devices are deployed in remote (\eg wildfire control) or other hard to access areas (\eg in the user's home).

Some IoT systems are \textit{close-loop} systems. These act based on sensors measurements in order to maintain a predictable output (feedback-loop). Examples are Cyber-Physical Systems (CPS) and some Industrial IoT systems~\cite{Bordel2017}. Other systems are \textit{open-loop}. These take input under consideration but do not react only based on those inputs (no feedback-loop)~\cite{Bloom2018}. As a result, making IoT \textit{open-loop} (there is no verification that an actuator performed the required operation) systems resilient is harder than \textit{closed-loop} ones, due to the lack of feedback. 

Nonetheless, any kind of IoT systems should be capable of reconfiguring themselves to recover from failures. A self-healing enabled system should be able to \textit{detect disruptions, diagnose the failure root cause and derive a remedy, and recover with a sound strategy} in a timely fashion~\cite{Psaier2011}.

The existing approaches for fault-tolerance (and \textit{self-healing}) typically follow (1)~\textit{reactive} methodology where errors are detected and then recovered from (\eg complex event processing, system watchdogs and supervisors),(2)~\textit{proactive} (also known as \textit{preventive}) methodology where errors are \textit{predicted} and avoided before faults being triggered using machine learning and other predictive mechanisms (\cf \pattern{Predictive Device Monitor}~\cite{Ramadas2017}), or, (3)~a combination of both~\cite{Psaier2011}.

Athreya \etal~\cite{Athreya2013} suggest devices should be able to manage themselves both in terms of configuration (self-configuration) and resource usage (self-optimization), proposing a measurement-based learning and adaptation framework that allows the system to adapt itself to changing system contexts and application demands. Although their work has some considerations about resilience to failures (\emph{e.g.}, power outages, attacks), it does not address self-healing concerns.

The concept of \emph{responsible objects}~\cite{Angarita2015}, states that \emph{things} should be self-aware of their context (passage of time, progress of execution and resource consumption), and apply \emph{smart} self-healing decisions taking into account component transaction properties (backward and forward recovery). Their approach shows limitations, \emph{viz.} (1) when applied to time-critical applications, as it is not clear how much time we should wait for a transaction to finish, (2) some processes, such as those triggered by emergencies, cannot be compensated, and (3) when is it acceptable to perform \emph{checkpoints} in a continuously running system that cannot be \emph{rolled-back}? It also ignores the nature of constrained devices (\emph{e.g.}, limited memory, power) that might challenge the implementation of transactions. 

Aktas \etal~\cite{Aktas2019} are amongst the first to purpose runtime verification mechanisms to identify issues by resorting to a complex event processing (CEP) technique and \emph{``applying rule‐based pattern detection on the events generated real-time''}. They do not address \emph{self-healing} and only convey a summary of problems or possible problems to human operators. Leotta \etal~\cite{Leotta2018} also present runtime verification as a testing approach by using UML state machine diagrams to specify the system's expected behaviour. However, their solution depends on the definition of a formal specification of the complete system, which is unfeasible for highly-dynamic IoT environments (\emph{e.g.}, dynamic network topology).

There was also some literature found on the evolution of Node-RED with fault-tolerance considerations. The work by Margarida \etal~\cite{Margarida2020} propose a modification to Node-RED that allows the automatic decomposition and partitioning of the system towards higher decentralization, by running a custom firmware on the IoT devices. The system reliability is increase since Node-RED nodes could be automatically assign nodes to devices based on pre-specified properties and priorities, thus when abnormal run-time conditions were observed, the system reconfigured itself (\ie self-configuration and self-healing). The work was validated by running a set of experimental scenarios on both virtual and physical devices. A similar work by Szydlo \etal~\cite{szydlo2017flow} decomposes Node-RED flows into code artifacts that allow a mostly seamless integration with IoT devices, however there is no automation of the initial flow's decomposition and partitioning, nor efforts in detecting bottlenecks or addressing their impact and taking measures to reduce the impact of such issues on service delivery.

Other works~\cite{Blackstock2014,Noor19,Cheng17} focus on creating a distributed version of Node-RED were the flows can be deployed across multiple Node-RED instances (\eg on-premises and in the cloud). This increases the resilience of the Node-RED itself by avoiding its out-of-the-box centralized design. However, no further considerations about the IoT devices themselves and their failures are taken into account.

To the best of our knowledge, no current work attempts to provide known self-healing strategies and mechanisms to \emph{visual} and other \emph{low-code} approaches for IoT development, namely, no runtime verification mechanisms for visual programming environments nor a straightforward way of visually-defining fallback measures in case of both intermittent and total failures of components. This is not unexpected, as Leotta \etal~\cite{Leotta2018} point out that \emph{``software testing (in IoT) has been mostly overlooked so far, both by research and industry,''} and later corroborated by Seeger \etal~\cite{seeger2019optimally}, claiming that most of the research being conducted in visual programming for IoT has been disregarding failure detection and recovery.

\section{Self-healing for IoT}
\label{sec:work}

\begin{table}[!htp]
\begin{minipage}[t]{\linewidth}
\caption{Identified error detection (probing) patterns.}
\setlength{\extrarowheight}{.15em}
\label{table:probes}
\begin{tabularx}{\linewidth}{@{}>{\RaggedRight}p{2.12cm}@{}X@{}}
\toprule
\textbf{Pattern} & \textbf{Description} \\ \midrule
\pattern{Action Audit} & Guarantee that required actions are triggered when need by checking its effects. \\
\pattern{Suitable Conditions} &  Check if surrounding conditions are suitable for device operation.  \\
\pattern{Reasonable Values} &  Check if the values fit into a reasonable pattern for the device and its operational constraints. \\
\pattern{Unimpaired Connectivity} & Check that the different parts of the system can communicate using the primary communication infrastructure. \\
\pattern{Within Reach} & Guarantee that mostly idle devices are able to communicate when needed by making routine trials.  \\
\pattern{Component Compliance} & Check if the system parts are running the software they should in the way they are expected to. \\
\pattern{Coherent Readings} & Compare sensor data from various sources to improve and ensure sensing data quality.  \\
\pattern{Internal Coherence} & Regularly check if the system internal state correctly mirrors the actual devices' state. \\
\pattern{Stable Timing} & Check if devices are sending messages at the expected periodicity. \\
\pattern{Unsurprising Activity} & Check if devices are producing a suspicious number of messages as that might indicate severe hardware or logic problem. \\
\pattern{Timeout} & Keep a timer running since the first action and observe if a reaction happened, otherwise a problem has happened. \\
\pattern{Conformant Values} & Check if the device readings are in conformance with the devices' manufacturer specification. \\
\pattern{Resource Monitor} &  Monitor the system resources continuously, checking if the resources suit the operational needs.\\
\bottomrule
\end{tabularx}
\end{minipage}
\begin{minipage}[b]{\linewidth}
\caption{Identified recovery and maintenance of health patterns.}
\setlength{\extrarowheight}{.15em}
\label{table:recovery}
\begin{tabularx}{\linewidth}{@{}>{\RaggedRight}p{2.12cm}@{}X@{}}
\toprule
\textbf{Pattern} & \textbf{Description} \\ \midrule
\pattern{Redundancy} & Use redundancy as a way of minimize the impact of a faulty part. \\
\pattern{Diversity} & Use different entities to achieve a common goal and reduce the impact of faulty parts.\\
\pattern{Runtime Adaptation} & Enable the system to use different infrastructure seamlessly during operation.\\
\pattern{Debounce} & Filter or aggregate events to meet operational timing constraints.\\
\pattern{Balancing} & Distribute software and load between available resources to meet operational demands.\\
\pattern{Compensate} & Ensure system operation by mitigating sensing errors by having mechanisms that can compensate missing or erroneous information.\\
\pattern{Timebox} & Only process an order in a specific period to respects the system operational constraints, filtering the remaining requests within a time-span.\\
\pattern{Checkpoint} & Preserve the current (or most recent) system state to avoid repeating actions or changing devices states to defaults in case of disruption.\\
\pattern{Reset} & Perform system resets periodically or when some error is detected as a preventive measure.\\
\pattern{Consensus Among Values} & Compare information from several sources enforcing a consensus before taking a decision.\\
\pattern{Circumvent and Isolate} & Circumvent and isolate failing parts, by disabling faulty components and reconfiguring the system to ignore them.\\
\pattern{Flash} & Restore a device to manufacturer settings with a trusted software version.\\
\pattern{Calibrate} & Ensure the accuracy of the data collected by (re)calibrating devices' to meet the expected behaviour.\\
\pattern{Rebuild Internal State} & Rebuild the internal state of the system to comply with (mirror) the current system state.\\
\bottomrule
\end{tabularx}
\end{minipage}
\end{table}

\subsection{Self-healing Patterns for IoT} \label{ssec:patterns-self}

Some of our previous work on dependable IoT systems focused on the systematization of widespread knowledge from both scientific and grey literature regarding fault-tolerance and dependability on various systems typologies including hardware-specific, industrial systems, software applications and cloud computing ~\cite{europlop2020}. This systematization presents 27 patterns split into two categories: (1) error detection patterns, and (2) recovery and maintenance of health patterns. In the first category, error detection, we list patterns that focus on checking the health of the system and its parts. Some patterns address cross-cutting issues to almost any other computing system but under the IoT point-of-view. In contrast, others focus on checking the correct operation of its sensing and actuating parts and are listed in Table~\ref{table:probes}. Regarding the second one, recovery and maintenance of health patterns, the patterns address the issue of guaranteeing normal system operation even when failures with different system parts occur. Self-healing is then achieved by contemplating strategies encompassing traditional fault-tolerance approaches such as redundancy, and mechanisms tailored for sensing and actuating errors and misbehaviour's, being listed in Table~\ref{table:recovery}. Some additional patterns presented in other works are also considered, even if they do not directly focus on enabling self-healing capabilities. Namely, the existence of a \pattern{Device Registry}~\cite{Ramadas2017,reinfurt2017internet} allows to check, during runtime, what resources are available, their capabilities and exposed services. Complementary patterns regarding device configuration (\ie self-configuration) are also considered, namely: \pattern{Automatic Client-Driven Registration} and \pattern{Automatic Server-Driven Registration}~\cite{reinfurt2017internet}.

\subsection{Self-Healing Extensions}

Taking as ground-work the previous proof-of-concept, which asserted the feasibility of our approach, altogether with the effort carried on the systematization of the existing knowledge of self-healing for IoT in the form of patterns (\cf Subsection~\ref{ssec:patterns-self}), a minimal working set of nodes was devised that enabled the implementation of self-healing behaviours with new, or by modifying already existent, Node-RED \textit{flows}. A total of 17 nodes were implemented, as presented in Table~\ref{table:shennodesmap}. These nodes do not cover all the presented patterns since (1) some patterns are not applicable within Node-RED, (\eg \pattern{Calibrate}) and (2) some patterns are enabled by Node-RED's built-in nodes or available extensions (\eg \pattern{Timeout}). The implementation of the JavaScript nodes was validated against their specification using software testing approaches (Node-RED's \textit{test-helper} framework).

\begin{table*}[!htbp]
\caption{Self-Healing Extensions node pallet and map to self-healing patterns}
\label{table:shennodesmap}
\setlength{\extrarowheight}{.25em}
\begin{tabularx}{\linewidth}{>{\RaggedRight}p{2.6cm} X >{\RaggedRight}p{4.2cm}}
\toprule
Node & Description & Enabled Patterns\\ 
\midrule
\texttt{action-audit} & After a trigger action is given, a sensor reading that acknowledges the action is waited for until a timeout occurs. & \pattern{Timeout}, \pattern{Action Audit}  \\ 
\texttt{balancing} & Distributes computation tasks (messages) among available resources (\eg similar or redundant devices), by distributing the messages among nodes using either \textit{Round Robin}, \textit{Weighted Round Robin} or \textit{Random} strategies. & \pattern{Balancing}, \pattern{Redundancy}  \\ 
\texttt{checkpoint} & Stores the last input message of a node, replaying it in case of Node-RED failure (with a time-to-live threshold). & \pattern{Checkpoint} \\ 
\texttt{compensate} & Compensate missing values using pre-defined strategies, complying with the expected values periodicity. Also provide  confidence analysis in consecutive compensations. & \pattern{Compensate} \\ 
\texttt{debounce} & Adjusts periodicity of messages to meet  target periodicity requirements (\eg actuator response capability), by operating as a rate-limit with aggregation/filtering capabilities. & \pattern{Stable Timing}, \pattern{Debounce}, \pattern{Timebox} \\ 
\texttt{flow-control} & Enable/disable \textit{flows}, allowing to adapt to changes/disruptions in the system. & \pattern{Circumvent and Isolate}, \pattern{Runtime Adaptation} \\ 
\texttt{heartbeat} & Heartbeat that check the alive status of system parts connected over HTTP and MQTT. & \pattern{Within Reach}, \pattern{Timeout}, \pattern{Unimpaired Connectivity}\\ 
\texttt{http-aware} & Periodically probes the network for running services (on specified ports), discovering new ones or checking if some has disappeared (\ie disconnected).  & \pattern{Within Reach}, \pattern{Unimpaired Connectivity}, \pattern{Device Registry}\\ 
\texttt{kalman-filter} & Provides an implementation of the Kalman noise filter~\cite{bishop2001kalman} which uses statistical predictors to reduce the effect of random noise on measurements.  & \pattern{Compensate}\\ 
\texttt{network-aware} & Periodically scan of the local network for finding new or disconnected devices and hosts (discovery). & \pattern{Within Reach}, \pattern{Unimpaired Connectivity}, \pattern{Device Registry}\\ 
\texttt{redundancy} &  Manage redundant instances of Node-RED, setting a new master instance on the case of disruption of a master instance and reconfigure in case of recovery. & \pattern{Redundancy}\\ 
\texttt{readings-watcher} & Check if sequential sensor readings are meaningful and correct by checking for minimum changes, maximum changes or stuck-at anomaly (same sequential reading). & \pattern{Reasonable Values} \\ 
\texttt{replication-voter} & Selects a value (message) taking into account several input messages (\eg array of sensor readings), based on a consensus (\eg majority). & \pattern{Redundancy}, \pattern{Diversity}\\ 
\texttt{resource-monitor} & Checks telemetry data reported by the different system parts against near-maximum (or near-minimum) thresholds. & \pattern{Resource Monitor}\\ 
\texttt{threshold-check} & Checks if measurements are within the operational specifications of the device. Can also be used to check if the surrounding conditions allow correct device operation. & \pattern{Reasonable Values}, \pattern{Suitable Conditions} \\ 
\texttt{timing-check} & Checks if the periodicity of incoming messages matches the expected rate. & \pattern{Unsurprising Activity} \\ 
\texttt{device-registry} & Maintains a registry within Node-RED with all devices in the system which adapts and triggers events as connected devices change.  & \pattern{Within Reach}, \pattern{Device Registry}\\
\bottomrule
\end{tabularx}
\end{table*}

\section{Experiments and Results}
\label{sec:experiments}

We performed experiments focusing on how different nodes could be combined to improve system dependability. These were run in a testbed presented in~\cite{selfheal20}: a \textit{SmartLab}. It consists in three sensing nodes, four actuators, and an on-premises server running Node-RED, an MQTT message broker, and a database. An additional host exists as a replica to ensure minimal redundancy of the Node-RED instance, providing a fallback mechanism (\cf \pattern{Redundancy}).

Failures were forced into the system --- by physically and virtually injecting faults --- to assert the system's behaviour, \ie its ability to delivery correct service when degradation occurred. These faults were injected arbitrarily during the time of observation of the different scenarios to closely mimic their real-world typical occurrence (\eg device malfunction, delayed communication, connectivity failures, power supply instabilities and resets due to unhandled exceptions).

\subsection{Sensor Failure Scenario}

\begin{figure*}[!h]
\centering
\includegraphics[width=.9\textwidth]{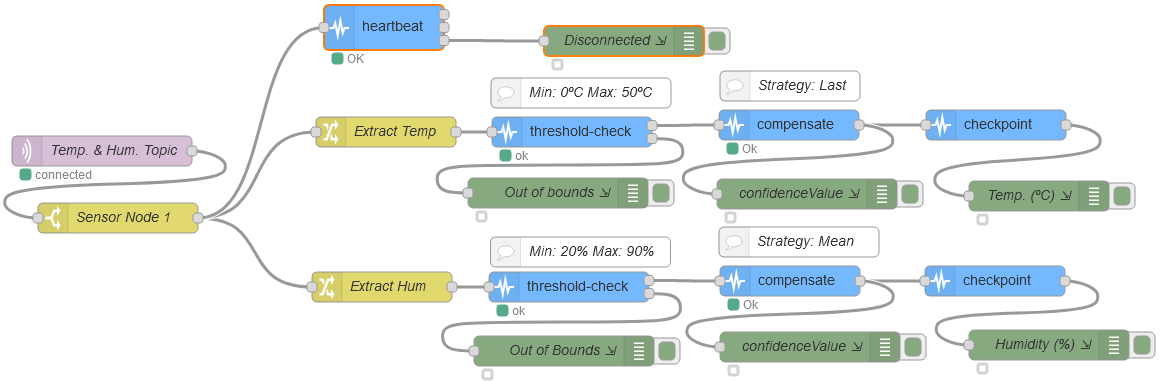}
\caption{Experiment with failure of sensing device and \pattern{Compensate} maintenance of health pattern, along with \pattern{heartbeat probe.}}
\label{fig:compensatechart-flow}
\end{figure*}

\begin{algorithm}
\Input{reading: $\mathbb{R}$}
\Output{$\langle$reading: $\mathbb{R}$, error$\rangle$\Comment*[r]{egress ignores `\_' messages}}
\BlankLine
\Init{config: $\{$\\
    \Indp 
    low: $\mathbb{R}$, high: $\mathbb{R}$\\
    \textbf{inv} low $\leq$ high\\
    \Indm$\}$\\
    }
\BlankLine
\OnInput{
    \If{config.low $\leq$ reading $\leq$ config.high}{\Return $\langle$reading, \_$\rangle$}
    \lElse{\Return $\langle$\_, error$\rangle$}
}
\caption{\texttt{threshold-check} node.}
\label{alg:threshold-checker}
\end{algorithm}

\begin{algorithm}[!h]
\Input{reading: $\alpha$}
\Output{reading: $\alpha$}
\BlankLine
\Init{config: $\{$\\
    \Indp
    historyMaxSize: $\mathbb{Z}_{>1}$, interval: $\mathbb{R}_{>0}$,\\
    strategy: $[\alpha] \rightarrow \alpha \gets$ s $\in \{$avg, max, min, last, ...$\}$\\
    \Indm$\}$\\
    timer $\gets$ $newTimer($config.interval$)$\\
    msgHistory: $[\alpha] \gets$ $[\:]$ 
    }
\BlankLine
\OnInput{
    \If{$|msgHistory|$ $\geq$ config.historyMaxSize}{
        $delete($msgHistory$_0)$
    }
    msgHistory $\gets$ msgHistory $+\!+$ reading\\
    timer.$start()$ \Comment*[r]{(Re)start timer}
    \Return reading
}
\BlankLine
\OnTimeout{
    reading $\gets$ config.$strategy($msgHistory$)$\Comment*[r]{Injects input}
    \Trigger onInput
}
\caption{\texttt{compensate} node.}
\label{alg:compensate-readings}
\end{algorithm}

\begin{algorithm}
\Input{message}
\Output{message}
\BlankLine
\Init{config: $\{$timeToLive: $\mathbb{R}_{>0}\}$\\
      store: $\{$timestamp, lastMessage$\}$\\
      timestamp $\gets$ store.timestamp \textbf{or} \textsc{nil}\\
      lastMessage $\gets$ store.lastMessage \textbf{or} \textsc{nil}\\
      \If{lastMessage $\neq$ \textsc{nil}}{
        aliveTime $\gets$ time.$now()$~$-$~store.timestamp\\
      \If{aliveTime $\leq$ config.timeToLive}{
            retained $\gets$ lastMessage\\
            lastMessage $\gets$ \textsc{nil}\\
            \Return retained
        }
      }
    }
\BlankLine
\OnInput{
    store.timestamp $\gets$ time.$now()$ \Comment*[r]{store is persistent}
    store.lastMessage $\gets$ message
    
    \Return message
}
\caption{\texttt{checkpoint} node.}
\label{alg:checkpoint}
\end{algorithm}

\begin{algorithm}
\Input{message}
\Output{$\langle$ping, ok, error$\rangle$\Comment*[r]{egress ignores `\_' messages}}
\BlankLine
\Init{config: $\{$\\
      \Indp
      ping: message, ok: message, error: message,\\
      mode $\gets$ m $\in \{$ passive, active $\}$, timeout: $\mathbb{R}_{>0}$\\
      \Indm$\}$\\
     timer $\gets$ $newTimer($config.timeout$)$\\
    }
\BlankLine
\OnInput{
    timer.$restart()$\\
    \lIf{config.mode = passive}{\Return $\langle$\_, config.ok, \_$\rangle$
    }
    \lElse{\Return $\langle$config.ping, config.ok, $\_\rangle$
    }
}
\BlankLine
\OnTimeout{
   timer.$restart()$\\
   \Return $\langle$\_, \_, config.error$\rangle$
}
\caption{\texttt{heartbeat} node.}
\label{alg:heartbeat}
\end{algorithm}

A sensing node (\textit{Sensor Node 1}) with a humidity and temperature sensor, connected over MQTT, producing values every 60 seconds, is connected to the Node-RED \textit{flow} depicted in Fig.~\ref{fig:compensatechart-flow}. The sensor is a DHT11, capable of reading temperatures in the range of $[0,50]$\degree C and humidity in the range $[20,90]\%$. In parallel, a passive \texttt{heartbeat} (\cf Algorithm~\ref{alg:heartbeat}) checks if the \textit{Sensor Node 1} fails to produce any message within a given interval, triggering an error accordingly:

\begin{enumerate}
    \item the \texttt{threshold-check} (\cf Algorithm~\ref{alg:threshold-checker}), verifies if both readings are within the expected values for the sensor, dropping values out-of-bounds;
    \item the \texttt{compensate} (\cf Algorithm~\ref{alg:compensate-readings}) verifies if, at some point, the sensor readings rate do not match the expected periodicity (\ie 60 seconds). If not matched, estimation is done, and a corresponding message is sent at the expected interval: (1) the last reading for the temperature, and (2) the mean of the last ten readings for humidity;
    \item the \texttt{checkpoint} (\cf Algorithm~\ref{alg:threshold-checker}) ensures that if there is any disruption that resets the Node-RED flow (\eg host reboot), the last message is re-sent (if within the message Time-to-live configured limit).
\end{enumerate}

\begin{figure}[h]
\centering
\includegraphics[width=\linewidth]{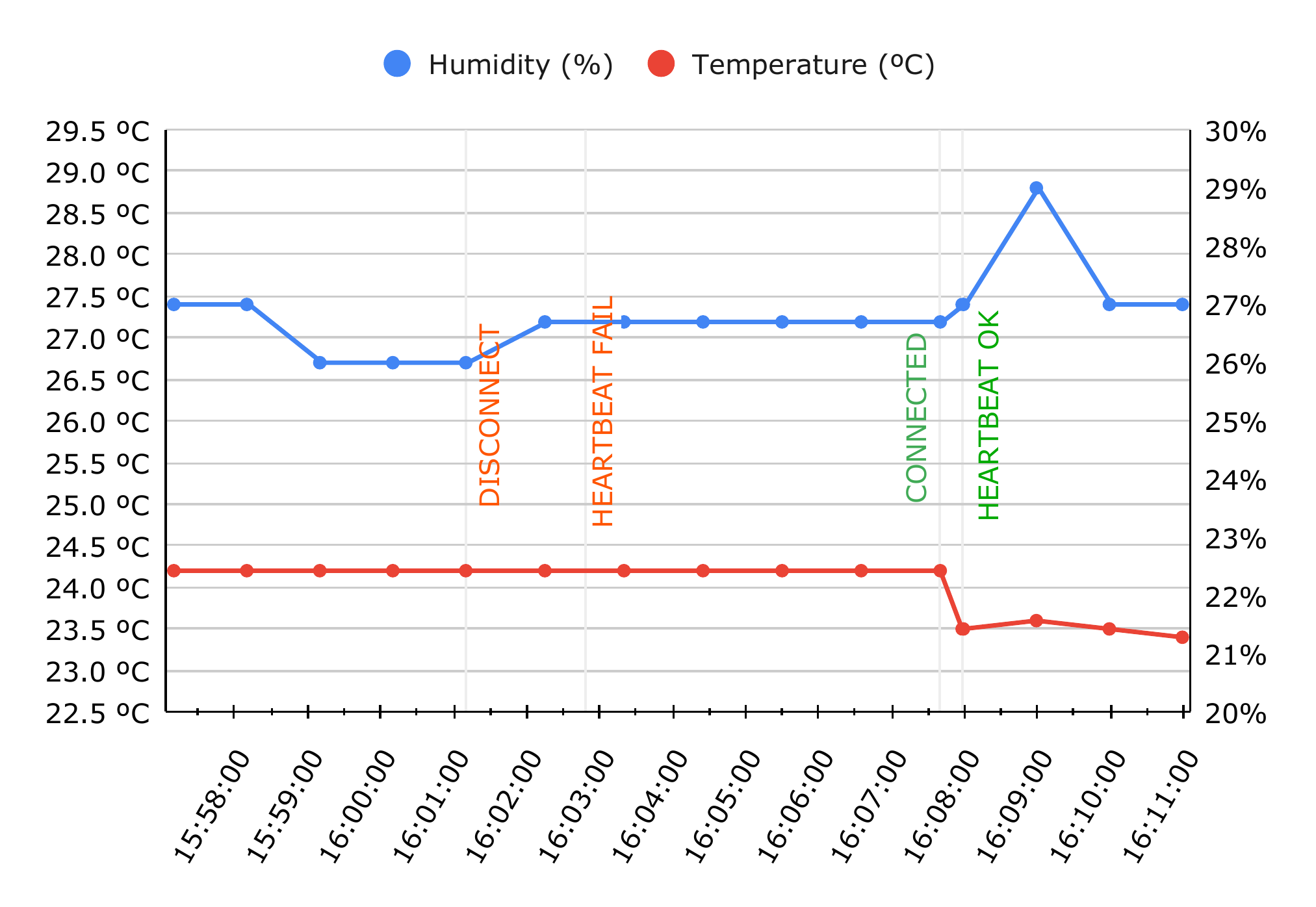}
\caption{Experiment with failure of sensing device and \pattern{Compensate} maintenance of health pattern, along with \pattern{heartbeat probe.}}
\label{fig:compensatechart}
\end{figure}

As the first experiment, a total failure of a sensing node was replicated by disconnecting the sensor node from power at a random moment (this can be seen in Fig.~\ref{fig:compensatechart}). Soon after the sensor node was disconnected, the heartbeat failed. As expected, the \texttt{compensate} node triggered and compensates the missing values using the configured strategies. When the device reconnects, the \texttt{compensate} node stopped producing values, and real readings are used. It is observable that when the device recovers there are two almost sequential readings, not matching the expected periodicity; this could be further managed --- if required by the receiver node or device --- using a \texttt{debounce} node to ensure that values are always at the same periodicity.

\subsection{Load Spike Scenario}

\begin{figure}[!ht]
\centering
\includegraphics[width=1\linewidth]{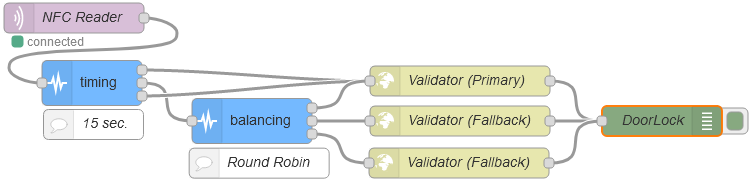}
\caption{Balancing the validation of identity cards (NFC) via an external service (\emph{e.g.} HTTP request) when a load spike happens (increased number of cards swiped per unit of time).}
\label{fig:balancing-sample}
\end{figure}

An access control device with an NFC reader is connected over MQTT to a Node-RED \textit{flow} depicted in Fig.~\ref{fig:balancing-sample}. The reader is placed at the entry point of the lab and ensures that every NFC card is validated using an external service, of which there are one primary host and two additional backup ones to be used in the case of exceptional usage spikes --- more than one card read in a 15s. window. The \textit{flow} in Fig.~\ref{fig:balancing-sample} ensures that all the card validation requests happen as fast as possible. There are a number of self-healing nodes in-place to assure this:

\begin{enumerate}
    \item the \texttt{timing-check} verifies the frequency at which the cards are being swiped in the NFC reader, categorizing (and splitting) them in accordance: \textit{too fast}, \textit{too slow} and \textit{normal} (using as reference the 15-second estimated time between readings); 
    \item the \texttt{balancing} node, which handles readings that are coming as \textit{too fast}, distributes them among the available \textit{validators}, ensuring a distribution in accordance to the configured strategy (\eg round-robin), thus reducing the load in the  primary host.
\end{enumerate}

\begin{figure}[!htb]
\centering
\includegraphics[width=\linewidth]{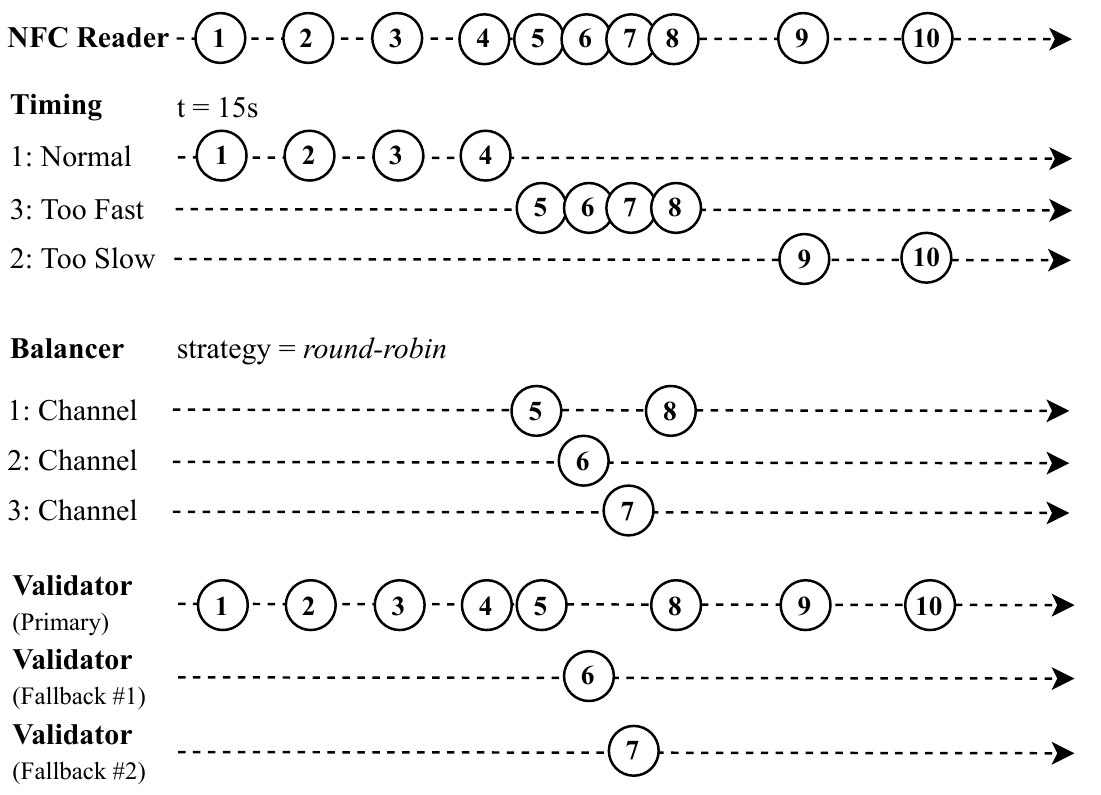}
\caption{Marble diagram of the messages between the different nodes (output messages) on the \textit{flow} of the load spike scenario, being the NFC reader the \textit{producer} of messages and the validators the final \textit{consumers}.}
\label{fig:marble_balancing}
\end{figure}

This behaviour is depicted in the marble diagram of Fig.~\ref{fig:marble_balancing}, replicating the behaviour recorded at the testbed.

\subsection{Redundancy Scenario}

Although \pattern{Redundancy} is one of the most common patterns found in fault-tolerant systems, having it implemented in Node-RED allows the definition of recovery behaviors using \emph{flows} that go beyond simply turning on or off a whole Node-RED runtime in the traditional redundant unit fashion.

\begin{figure}[!ht]
\centering
\includegraphics[width=\linewidth]{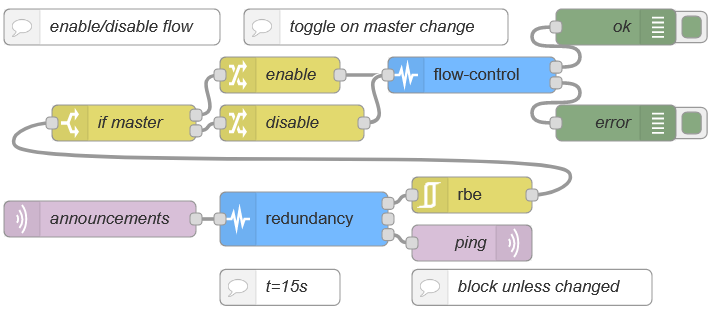}
\caption{Managing two Node-RED instances (\texttt{redundancy}), and adapting the behaviour of the system (\texttt{flow-control}) when the master instance changes\protect\footnotemark. The \texttt{redundancy} node is configured with a \pattern{timeout} that triggers a new election when a redundant instance stops \textit{pinging} for 15 seconds. The \texttt{rbe} (report-by-exception) node is part of the default palette and deduplicates sequentially repeated messages.}
\label{fig:redundancy-flow}
\end{figure}

There are two instances of Node-RED running at different hosts, with a common \textit{flow} that carries a common task: (1)~receiving sensor data from \textit{Sensor-Node-1} over MQTT (with a frequency of 1 reading per minute), (2)~extracting the temperature, (3)~asserting the validity of the data (\texttt{threshold-check} and \texttt{readings-watcher}), and (4)~posting the data to an external service (an HTTP endpoint). The \textit{flow} depicted in Fig.~\ref{fig:redundancy-flow} is deployed in both instances, running a consensus algorithm\footnote{The consensus algorithm implemented select as a master instance the node with the highest last octet of the IP address.} to define a new master if the previously defined master instance fails. Both Node-RED instances are running simultaneously, optimizing the mean time to recovery (MTTR) after a Node-RED instance crash (\emph{a.k.a.} active-standby). However, the \textit{flow} is only active in one of them (mutually-exclusive). When the master instance crashes, an election occurs to determine a new master (effective until the old master instance recovers --- if it recovers).

\begin{figure}[htb!]
\centering
\includegraphics[width=\linewidth]{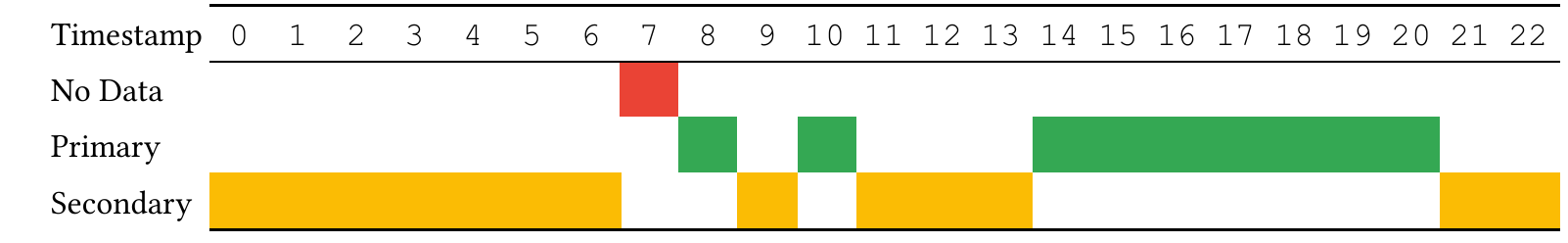}
\caption{Timeline (mins) depicting if the reading was provided by the primary (master) instance, secondary (fallback), or if no data arrived when expected.}
\label{fig:redundancy-heat}
\end{figure}

In this experiment, a new election occurs every 15 seconds (configuration of the \texttt{redundancy} node). The default master instance (the one with the highest octet) was turned off multiple times (randomly). The time was measured between the disconnection of the master instance until the sensor reading \textit{flow} resumes on the fallback instance (a \textit{ping} is done to an external service). A total of 10 measurements were made, and the MTTR of the system was $13.7s ~(\sigma=1.77s)$. 

As observed in Fig.~\ref{fig:redundancy-heat}, almost all sensor readings reach the external endpoint, even with the continuous toggling of the master Node-RED instance. During a test of 22 minutes, which should have resulted in 22 sensor readings, only one was lost.

\subsection{Threats to Validity}

The experiments were carried on a representative testbed deployed in a laboratory. While using a physical deployed testbed resembles a real-world scenario and can provide more realistic data and behaviours --- specially when a fault occurs --- when compared to simulated experiments, it also (1)~limits the number of devices used during the experiments due to additional costs, (2)~capturing failures-over-time requires long-running experiments, and (3)~the users that typically interact with the system have a level of expertise uncommon in most application domains. An inadequate selection of scenarios is also a threat to this work, since they have been hand-picked with prior knowledge of system and the SHEN implementation details. This can result in a bias in the selection, favoring issues that we have knowledge about and have more confidence that our solution will handle correctly, instead of the ones that are mostly like to occur. We attempt to mitigate this by mimicking real-world use cases; there certainly exists an opportunity for the creation of widely available datasets of IoT faults so authors can more robustly compare their solutions. Faults and implementation quirks of the underlying infrastructure (\ie Node-RED), might also influence the outcome of our experiments, so they are a confounding variable. We believe this has been mitigated by careful analysis of the expected outcome (\eg by observing the Node-RED's operational logs), though it is something to be aware of. 

\section{Conclusion}
\label{sec:conclude}

Built on top of previous published works, we have presented a solution that enables Node-RED users to improve overall system dependability via self-healing mechanisms. We have carried experiments using representative scenarios implemented on the \textit{SmartLab} testbed, showcasing the feasibility and effectiveness of the approach in terms of error detection, failure recovery and overall capability on health maintenance. 

We have also identified some limitations, which might be presented as future work, including: (1) resilience to network partitions, as the \texttt{Redundancy} node has no way of finding if there is already a master in the network; (2) most of the nodes still do not support the definition of reasonable margins (\eg in nodes that deal with timing constrains, a minor delay should be ignored instead of triggering the recover or maintenance action); (3) although its need was identified, no mechanism to synchronize the current system state between different Node-RED instances has been provided, and (4) the capabilities of the \pattern{Device Registry} pattern and device/service discovery are only partial due to the high heterogeneity and lack of standard of IoT systems. Further, current state-of-the-art do not present out-of-the-box solutions for distribution of computational tasks across devices in the heterogeneous IoT system beyond limited (both in scope or supported devices) proofs-of-concept~\cite{Margarida2020,Pinto18,szydlo2017flow,Blackstock2014,Noor19,Cheng17}, solutions which, if available, would offer foundations for other kinds of fault-tolerance mechanisms beyond the ones presented. Additionally, to be able to validate the correct functioning of our approach, there is the need of a solution that allows one to deliberately provoke failures in the system and report observed behaviours --- in cloud computing defined as \textit{chaos engineering} or \textit{fault injection}. 

\section*{Data Availability}

The extensions presented in this work are open-source and available on GitHub~\cite{gitrepo}, and they are ready to be installed using Node-RED extension manager (using the \texttt{npm} JavaScript package manager). A replication package is also available on Zenodo~\cite{jp_2021_4448164}.

\section*{Acknowledgments}

This work was partially funded by the Portuguese Foundation for Science and Technology (FCT), under the research grant {SFRH/BD/144612/2019}.

\bibliographystyle{IEEEtran}
\bibliography{refs}

\end{document}